# Optimization design of a micro-perforated panel absorber with 8.6 octave bands

Wang Xiaoming[1]    Liang Chen[1]    Mei Yulin[*2]

(1.School of Mechanical Engineering, Dalian University of Technology, Dalian 116024，China
2.School of Automotive Engineering, Dalian University of Technology, Dalian 116024，China )
meiyulin@dlut.edu.cn

**Abstract**：In order to improve low-frequency characteristics of micro-perforated panel absorbers, sound absorption structures composed of micro-perforated panels and expansion chambers are design, and an optimization design method is constructed based on the transfer function model and the simulated annealing algorithm. First, a single-chamber structure composed of a micro-perforated panel and an expansion chamber is build, and the sound absorption curve is simulated by the finite element method. Second, for the sake of enlarging the continuous absorption bandwidth with absorption coefficients not less than 0.8, a three-chamber structure is designed, which has a sound absorption bandwidth of 1277Hz (27-1304Hz) covering 5.6 octave bands. Then, the transfer function model of the structure is established, and a series of theoretical formulae are derived to calculate the absorption coefficients. Subsequently, the sound absorption bandwidths calculated by the theoretical formulae and the finite element method are compared, and the relative error is 3.68%. Finally, an optimization design method is constructed by combining the transfer function model and the simulated annealing algorithm, where the optimization objective is to maximize the absorption bandwidth and the optimization variables are structural parameters of the three-chamber structure. The results show, after optimization, the three-chamber structure exhibits an excellent sound absorption performance, with a continuous bandwidth of 1591Hz (4-1595Hz), realizing 8.6 octave bands.
**Keywords**：micro-perforated panel; sound absorption coefficient; transfer matrix; simulated annealing

## 1 Introduction

Noise is known as the invisible killer of environmental pollution (Meng, 2007). Compared with high-frequency noise, low-frequency noise has the characteristic of strong penetration. It has been a challenge for researchers to design sound structures with low-frequency broadband performance and with small dimensions suitable for engineering applications.

Micro-perforated panel (MPP) absorber was proposed by Maa in 1975, nearly 50 years ago (Maa, 1975). Although the MPP absorber has good sound absorption performance and has been widely used in the field of vibration and noise reduction, its low-frequency performance has been a shortcoming and has received many scholars' attentions. Yang et al. designed and optimized a multilayer MPP structure to absorb 100-500Hz sound waves (Yang et al., 2021). To expand the sound absorption band of the conventional MPP absorber, Liu et al. proposed different structure types based on the series-parallel coupling mechanism of the MPP (Liu et al., 2021). Rafique et al. combined the inhomogeneous MPP with the expansion chamber to construct a composite structure to improve transmission losses of low- and medium-frequency sound waves (Rafique et al., 2022). Wang et al. analyzed sound absorption performances of multi-chamber MPP structure by numerical simulation and experimental measurements, and designed an structure with absorption coefficient greater than 0.82 in the 340-2000Hz frequency band (Wang J. et al., 2023).Some scholars also constructed corrugated MPP absorber by transforming the traditional flat MPP into a corrugated MPP to improve the absorber's low frequency performance and sensitivity to the direction of sound waves (Qiu et al., 2023; Yang et al., 2023;Wang et al., 2020).

Finite element method and transfer matrix method have been widely used to design and analyze sound absorbers. Transfer matrix method is a theoretical method based on mathematical models and is suitable for low-frequency performance analysis. Compared with finite element method, transfer matrix method is more efficient, especially while repeatedly analyzing a complicated sound structure. Zaw et al. constructed a sound absorber through combining an expansion chamber and a Helmholtz resonator, and adopted transfer matrix method to analyze the effect of the parameters of the Helmholtz resonator on transmission losses (Zaw et al., 2018). Hu et al. built the transfer function model of a MPP with multiple perforation rates, and compared its sound absorption performance with that of a conventional single-layer MPP (Hu et al., 2020). Yan et al. designed a double-layer honeycomb MPP structure with adjustable rear cavity, and applied transfer matrix method to calculate its acoustic impedance (Yan et al., 2021).

For the sake of improving low-frequency characteristics of MPP absorbers, this paper designs a kind of sound structure composed of expansion chambers and MPPs, and adopts transfer matrix method and finite element method to analyze the sound absorption performances. Meanwhile, an optimization design method is constructed through combining transfer matrix method and simulated annealing algorithm to extend effective sound absorption bandwidth, and a sound absorption structure with 8.6 octave bands is obtained.

## 2 Structure design and simulation
### 2.1 The single-chamber structure

A single-chamber structure (Mei et al., 2021) is designed, as shown in Fig 1. It is composed of a main pipe, a MPP embedded inside the main pipe, and a cylindrical expansion cavity connected to the main pipe. The diameter and length of the main pipe are $d_m=10mm$, $l_m=100mm$. The MPP is placed at the entrance of the main pipe, and its perforation rate, thickness and aperture are $\sigma_h=0.025$, $t_h=0.6mm$, $d_h = 0.2mm$. The expansion chamber is placed at the end of the main pipe, and its diameter and thickness are $d_e = 60mm$, $t_e = 10mm$.

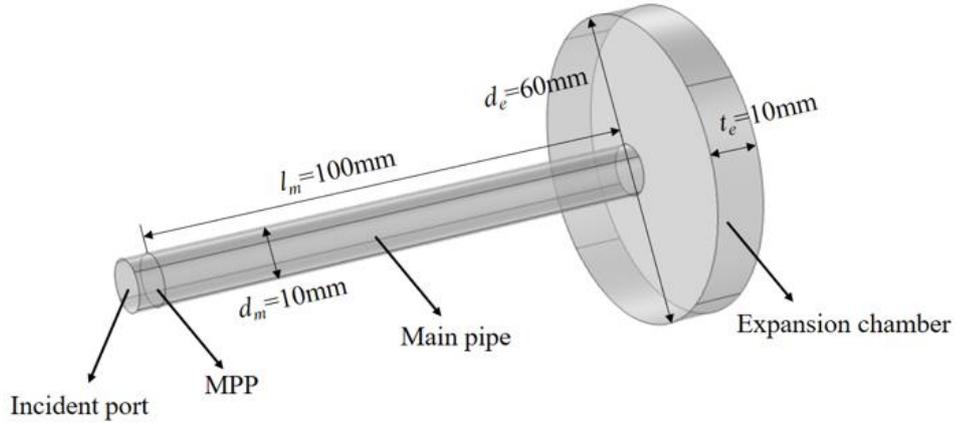

Fig.1 A single-chamber structure

COMSOL is used to build the finite element model of the single-chamber structure and to simulate its sound absorption performance. The plane wave with the sound pressure of 1Pa is loaded at the left end of the model, and the whole model is set as a hard sound field boundary. Set the working temperature and the sound velocity as 20°C and 343m/s. Define the effective sound absorption bandwidth as a frequency range with the sound absorption coefficients not less than 0.8.

In Fig.3, the green dashed line is the sound absorption coefficient curve simulated by COMSOL. It can be seen, in the low frequency range, the single-chamber structure has a continuous effective absorption bandwidth of 380Hz (97-477Hz) covering 2.3 octave bands, showing a good low-frequency sound absorption performance.

**2.2 The three-chamber structure**

In order to extend the effective absorption bandwidth in the low frequency range, a three-chamber structure (Wang and Mei, 2023) is designed by adding two expansion chambers and two MPPs, as shown in Fig.2. MPP 1 is still located at the entrance of the main pipe. MPP 2 is inserted between MPP 1 and the first expansion chamber, and its distance from the entrance of the main pipe is 98mm. MPP 3 is inserted between the first two expansion chambers, and its distance from the right end face of expansion chamber 1 is 10mm.

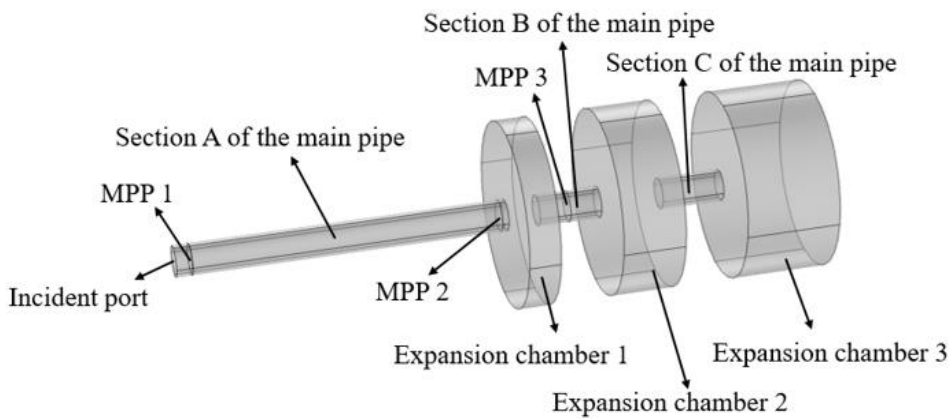

Fig.2 A three-chamber structure

Structural parameters of the three-chamber structure are listed in Table 1 and Table 2. The main pipe is divided into three sections with the lengths of $l_m^A$, $l_m^B$, $l_m^C$. For the *ith* expansion chamber, the corresponding diameter and thickness are $t_e^i$ and $d_e^i$, where $i=1,2,3$. For the *ith* MPP, the corresponding perforation rate, thickness and aperture are $\sigma_h^i$, $t_h^i$, $d_h^i$, where $i=1,2,3$.

Table.1 Structural parameters of the main pipe and the expansion chambers(mm)

| $d_m$ | $l_m^A$ | $l_m^B$ | $l_m^C$ | $t_e^1$ | $d_e^1$ | $t_e^2$ | $d_e^2$ | $t_e^3$ | $d_e^3$ |
|---|---|---|---|---|---|---|---|---|---|
| 10 | 100 | 20 | 20 | 10 | 60 | 20 | 60 | 30 | 60 |

Table.2 Structural parameters of MPPs

| $t_h^1$/mm | $d_h^1$/mm | $t_h^2$/mm | $d_h^2$/mm | $t_h^3$/mm | $d_h^3$/mm | $\sigma_h^1=\sigma_h^2=\sigma_h^3$ |
|---|---|---|---|---|---|---|
| 0.6 | 0.2 | 0.6 | 0.2 | 0.8 | 0.4 | 0.025 |

COMSOL is used to simulate the sound absorption performance of the three-chamber structure, and the results are shown in Fig.3, corresponding to the blue solid line. Through comparing the two absorption coefficient curves in Fig.3, it can be found, that the sound absorption performance of the three-chamber structure is improved significantly. The three-chamber structure achieves a continuous effective absorption bandwidth of 1277 Hz (27-1304 Hz) covering 5.6 octave bands, and its average absorption coefficient reaches to 0.87 in the effective absorption band.

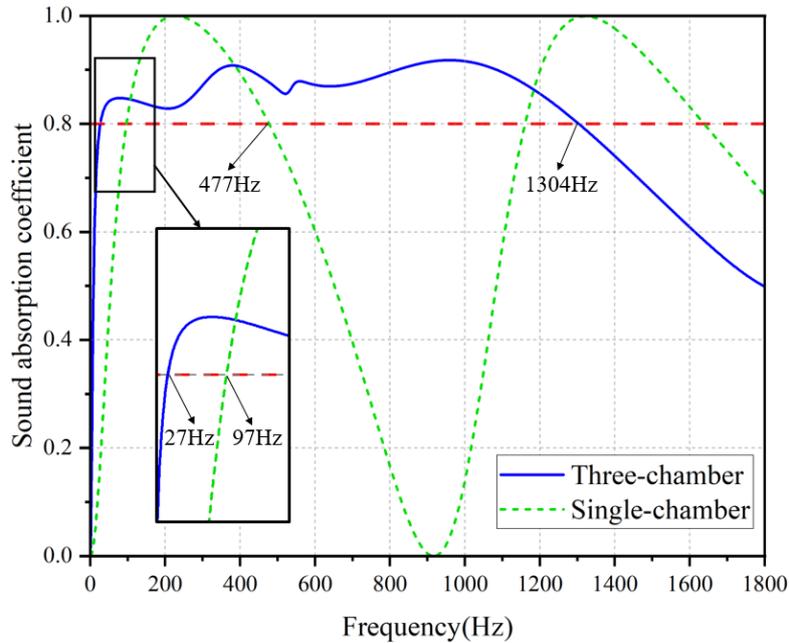

Fig.3 Sound absorption coefficient curves (COMSOL)

Fig. 4 illustrates the sound pressure cloud, at different frequencies, of the single-chamber and three-chamber structures. The sound absorption coefficients of the single-chamber structure are 0.99 at 228Hz, 0.8 at 97Hz and 0.5 at 55Hz, and the corresponding variation in the sound absorption performance is reflected in Fig. 4(a-c). It is obvious that the sound pressure inside the structure is almost 0Pa at 228Hz, and the sound pressures become bigger with the reduce in frequency from 97Hz to 55Hz. The sound absorption coefficients of the three-chamber structure are 0.91 at 380Hz, 0.8 at 27Hz and 0.5 at 9Hz, as shown in Fig. 4(d-f). The three-chamber structure exhibits better low-frequency absorption performance, such as at 27Hz, where sound waves propagating inside the structure are almost completely absorbed before the second expansion chamber.

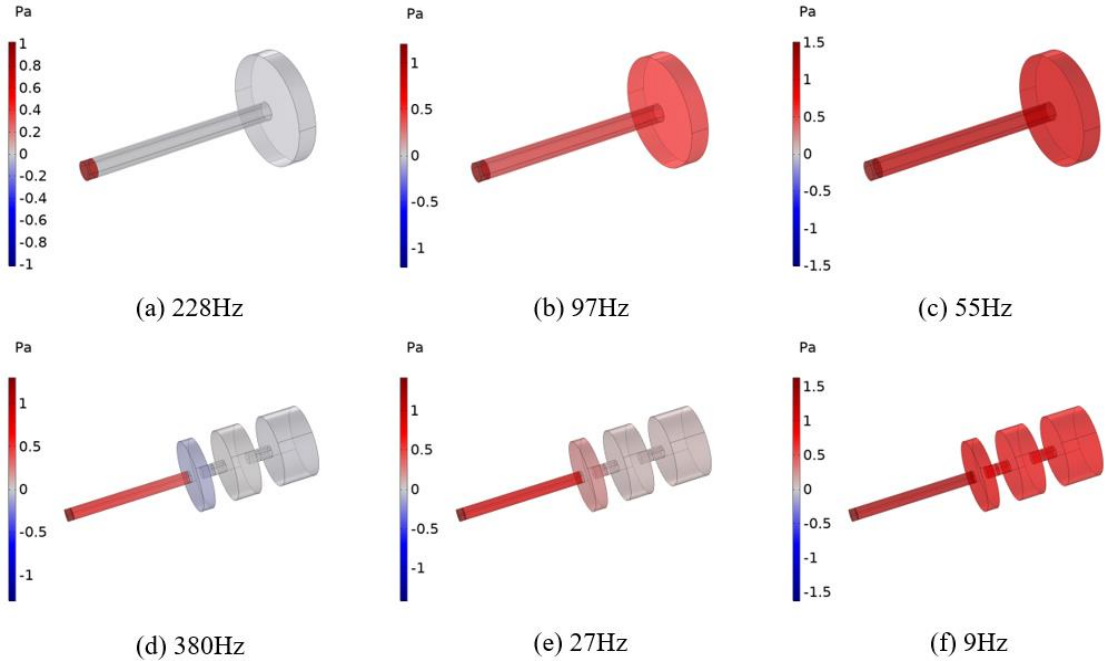

Fig.4 Sound pressure cloud map

## 3 Theoretical model

The three-chamber structure shows amazing sound absorption performance covering big octave bands. In order to verify COMSOL results and to further optimize the structure so as to realize bigger octave bands, transfer matrix method is used to establish the theoretical model of the three-chamber structure.

### 3.1 Transfer function model

Under the assumption of the plane wave, neglecting the variation of sound wave with time, the lumped parameters of a sound structure satisfy

$$\begin{pmatrix} P_z \\ U_z \end{pmatrix} = \begin{pmatrix} a_{11} & a_{12} \\ a_{21} & a_{22} \end{pmatrix} \begin{pmatrix} P_t \\ U_t \end{pmatrix} = A \begin{pmatrix} P_t \\ U_t \end{pmatrix} \quad (1)$$

where, $P_z$ and $U_z$ stand for the total sound pressure and total volume velocity at the input end; $P_t$ and $U_t$ denote the total sound pressure and total volume velocity at the output end; $A$ is the transfer matrix of the sound structure, describing the relationship between the input and output sound waves.

For a sound structure comprising $n$ sound elements connected in series, its transfer matrix can be expressed as

$$A = \begin{pmatrix} a_{11} & a_{12} \\ a_{21} & a_{22} \end{pmatrix} = \prod_{i=1}^{n} A_i \tag{2}$$

where, $A_i$ is the transfer matrix of the $ith$ sound element, $i = 1, 2 \cdots n$.

The three-chamber structure in Fig.2 comprises four types of sound elements, as shown in Fig.5.

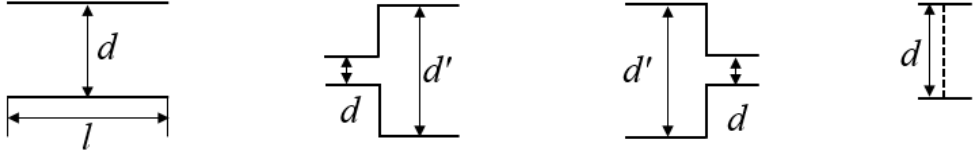

(a) Straight pipe element   (b) Expansion element   (c) Contraction element   (d) MPP element

Fig.5 Sound elements

For the straight pipe element in Fig.5(a), the transfer matrix can be expressed as (Zaw et al., 2018)

$$A_{Str} = \begin{pmatrix} \cos(kl) & jS \cdot \sin(kl) \\ \dfrac{j}{S} \sin(kl) & \cos(kl) \end{pmatrix} \tag{3}$$

where, $k$ is the wave number; $j = \sqrt{-1}$; $S = \pi(d/2)^2$.

The expansion element in Fig. 5(b) and the contraction element in Fig. 5(c) have the same transfer matrix, which can be written as

$$A_{Exp} = A_{Con} = \begin{pmatrix} 1 & 0 \\ 0 & 1 \end{pmatrix} \tag{4}$$

For the MPP element Fig. 5(d), the transfer matrix is

$$A_{MPP} = \begin{pmatrix} 1 & \dfrac{\rho_0 c_0}{S'} Z_{MPP} \\ 0 & 1 \end{pmatrix} \tag{5}$$

where, $\rho_0$ is the air density; $c_0$ the sound velocity in air; $S'$ the cross-sectional area of the pipe where the MPP are embedded; $Z_{MPP}$ the MPP impedance, satisfying (Maa, 1975)

$$Z_{MPP} = \frac{32\eta t_h}{\sigma_h \rho_0 c_0 d_h^2}\left[\left(1+\frac{K^2}{32}\right)^{\frac{1}{2}} + \frac{\sqrt{2}}{32}K\frac{d_h}{t_h}\right] + i\frac{\omega t}{\sigma_h c_0}\left[1+\left(9+\frac{K^2}{2}\right)^{-\frac{1}{2}} + 0.85\frac{d_h}{t_h}\right] \quad (6)$$

where, $t_h$, $d_h$, $\sigma_h$ are the MPP thickness, aperture and perforation rate; $\eta$ the air viscosity coefficient; $\omega$ the angular frequency; $K = d_h\sqrt{(\omega\rho_0)/(4\eta)}$.

The three-chamber structure can be divided into sixteen basic sound elements, as shown in Fig.6, including three MPP elements, three expansion elements, two contraction elements and eight straight pipe elements. Through applying the expressions (2)-(5), the transfer matrix A of the three-chamber structure can be figured out.

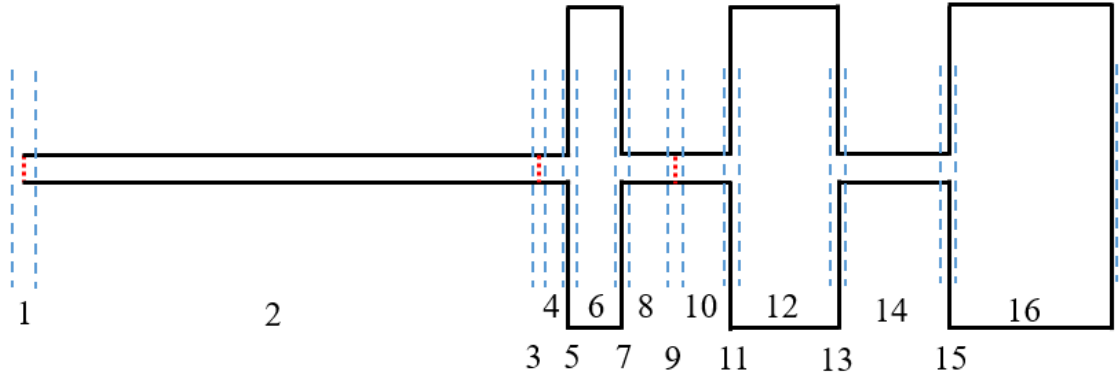

Fig.6 Distribution of sound elements

### 3.2 Sound absorption coefficients

The sound propagation in the sound structure is shown in Fig.7, where $P_i$ and $U_i$ are the sound pressure and the volume velocity of the incident wave, $P_r$ and $U_r$ are the sound pressure and the volume velocity of the reflected wave.

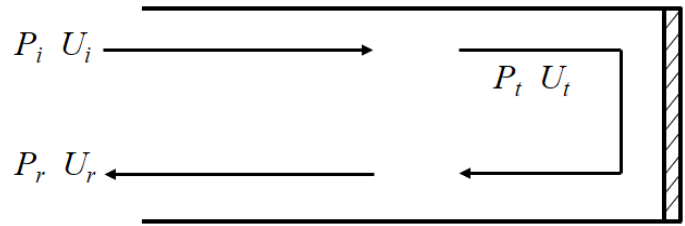

Fig.7 Sound propagation schematic diagram

While the sound wave propagates to the rigid wall, the output sound pressure $P_t$ is constant and the output volume velocity $U_t = 0\, m^3/s$. In this way, the expression (1)

can be rewritten as

$$\begin{pmatrix} P_z \\ U_z \end{pmatrix} = \begin{pmatrix} a_{11} & a_{12} \\ a_{21} & a_{22} \end{pmatrix} \begin{pmatrix} P_t \\ 0 \end{pmatrix} \tag{7}$$

Through solving the system of equations, the impedance of the sound structure can be calculated by

$$Z = \frac{P_z}{U_z} = \frac{a_{11}}{a_{21}} \tag{8}$$

At the input end of the sound structure, the sound pressure $P_z$ and the volume velocity $U_z$ satisfy

$$\begin{cases} P_z = P_i + P_r \\ U_z = U_i + U_r \end{cases} \tag{9}$$

Using $S_m$ to denote the cross-sectional area of the main pipe of the sound structure, the characteristic impedance of the sound structure is

$$Z_0 = \frac{\rho_0 c_0}{S_m} \tag{10}$$

For the incident and reflected sound waves, the ratio of sound pressure and volume velocity satisfies

$$\frac{P_i}{U_i} = -\frac{P_r}{U_r} = Z_0 \tag{11}$$

Then, the reflection coefficient of the sound structure can be expressed as

$$\Gamma = \frac{P_r}{P_i} = \frac{a_{11} - Z_0 a_{21}}{a_{11} + Z_0 a_{21}} \tag{12}$$

And the absorption coefficient of the sound structure can be obtained by (Lou et al., 2022)

$$\alpha = 1 - |\Gamma|^2 = 1 - \left| \frac{a_{11} - Z_0 a_{21}}{a_{11} + Z_0 a_{21}} \right|^2 \tag{13}$$

### 3.3 Theoretical prediction

Based on the transfer function model, the MATLAB program is developed to predict the absorption coefficient curve of the three-chamber structure, as shown in Fig.8. According to the theoretical prediction, the effective absorption bandwidth is 1324Hz (20-1344Hz), with a relative error of 3.68% compared with COSMOL results. In Fig.8, the theoretical prediction is well coincident with COSMOL curve, which provides a theoretical basis for the optimization design of the sound structure.

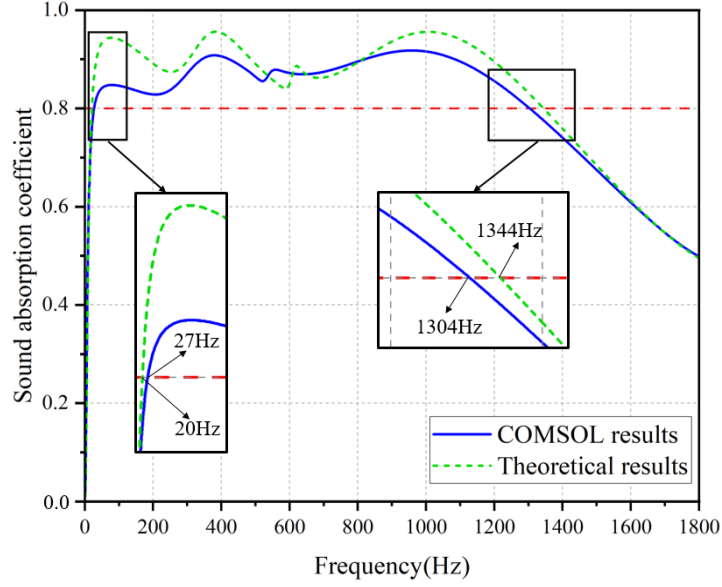

Fig.8 Sound absorption coefficient curves (the three-chamber structure before optimization)

## 4 Structure optimization

Based on the established theoretical model, the simulated annealing algorithm is adopted to construct the optimization design method to further broaden the effective sound absorption band of the three-chamber structure.

### 4.1 Optimization method

The simulated annealing algorithm was first proposed by Metropolis (Chiu, 2013), belonging to the stochastic optimization-seeking algorithm. Through combing theoretical model and the simulated annealing algorithm, the optimization design method of the three-chamber structure is constructed, which can be expressed as

$$\begin{cases} \text{Max} & f_h - f_l \\ s.t. & \min(\alpha) \geq 0.8 \\ & d_m \in [5,11] \\ & d_2 \in [40,70], \quad d_4 \in [50,80] \quad d_6 \in [50,100] \\ & l_1 \in [60,80], \quad l_1' \in [10,30], \quad l_2 \in [4,12] \\ & l_3 \in [4,10], \quad l_3' \in [4,10], \quad l_4 \in [10,30] \\ & l_5 \in [10,30], \quad l_6 \in [20,40] \end{cases} \quad (14)$$

The optimization objective is to maximize the effective sound absorption bandwidth, where $f_l$ and $f_h$ denote the low and high cutoff frequencies, respectively. The constraint is the effective sound absorption coefficients $\alpha \geq 0.8$. The optimization variables are the structural parameters of the three-chamber structure, as shown in Fig.9. The design domains are determined after considering spatial constraints and numerical simulation results.

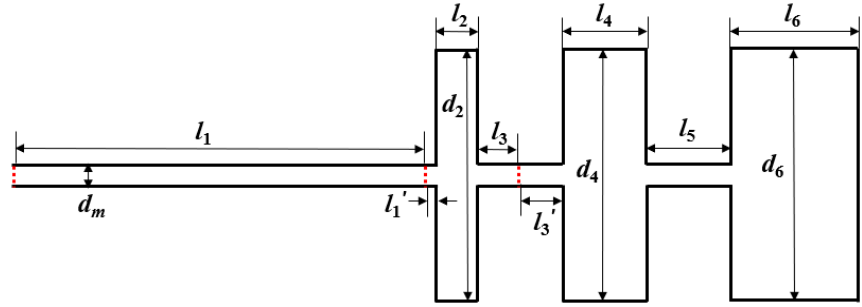

Fig.9 Optimization variables of the three-chamber structure

## 4.2 Optimization results

An iterative design is implemented through making use of the optimization method (14). Initial values of the optimization variables are given in Table 3. Set the initial temperature to be 100°C, the number of isothermal iterations to be 100, the temperature drop rate to be 0.2, and the termination temperature to be $10^{-6}$°C.

Table.3 Initial values(mm)

| $d_m$ | $d_2$ | $d_4$ | $d_6$ | $l_1$ | $l_1'$ | $l_2$ | $l_3$ | $l_3'$ | $l_4$ | $l_5$ | $l_6$ |
|---|---|---|---|---|---|---|---|---|---|---|---|
| 10 | 60 | 60 | 60 | 98 | 2 | 10 | 10 | 10 | 20 | 20 | 30 |

Table 4 lists optimization results of the structure parameters, and Fig.10 illustrates the sound absorption coefficient curves, predicted by the theoretical model, before and after optimization. According to the theoretical prediction, after optimization, the effective sound absorption bandwidth extends from 1324Hz (20-1344Hz) to 1591Hz (4-1595Hz) covering 8.6 octave bands, about 141% of the octave bands before optimization.

Table.4 Optimization results(mm)

| $d_m$ | $d_2$ | $d_4$ | $d_6$ | $l_1$ | $l_1'$ | $l_2$ | $l_3$ | $l_3'$ | $l_4$ | $l_5$ | $l_6$ |
|---|---|---|---|---|---|---|---|---|---|---|---|
| 5.6 | 41 | 57 | 97.8 | 69.6 | 10.4 | 8.9 | 4 | 9.3 | 18 | 21.2 | 36.9 |

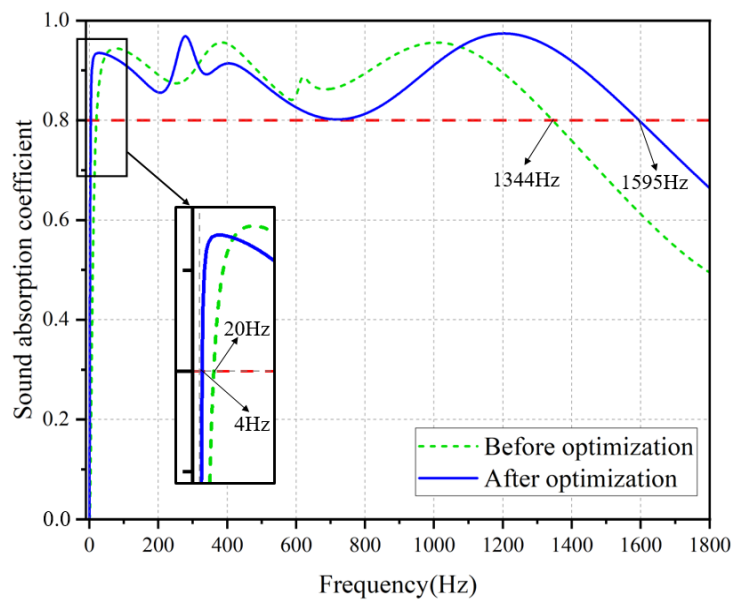

Fig.10 Sound absorption coefficient curves (Theoretical results of the three-chamber structure)

Fig.11. compares the sound absorption coefficient curves, calculated by the theoretical model and COMSOL, after optimization. According to COMSOL results, after optimization, the effective sound absorption bandwidth is 1576Hz (6-1582Hz) with a relative error of 0.94% compared with the theoretical prediction.

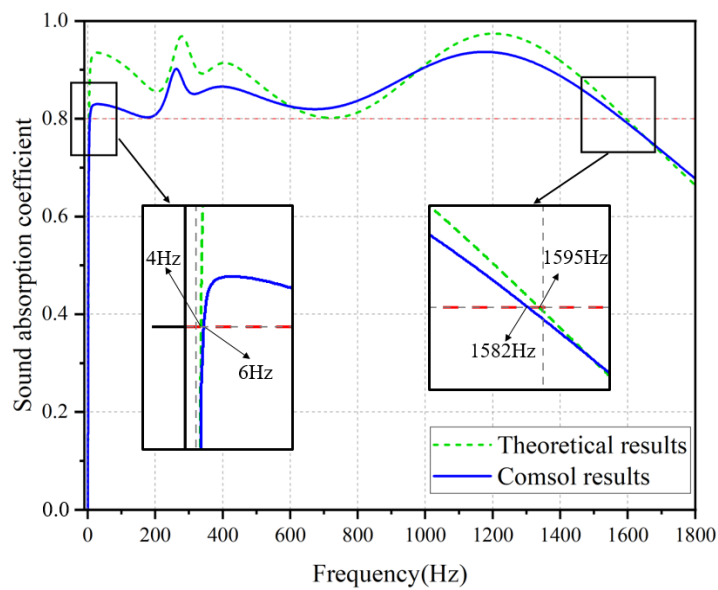

Fig.11 Sound absorption coefficient curves (the three-chamber structure after optimization)

Fig.12 shows the sound pressure cloud maps, calculated by COMSOL, of the optimized three-chamber structure. The sound absorption performances at different frequencies can be clearly observed. The sound absorption coefficient is 0.9 at 262Hz, corresponding to the peak of the blue solid line in Fig.11. It can be seen from Fig.12(a) that the sound waves are almost completely absorbed before propagating to the first expansion chamber. The sound absorption coefficient is 0.8 at 6Hz, corresponding to the low cutoff frequency of the blue solid line in Fig.11. It can be seen from Fig.12(b) that a large amount of sound energy is absorbed before propagating to the second expansion chamber. The sound absorption coefficient is 0.6 at 2Hz, and Fig.12(c) reflects the high sound pressure in the structure.

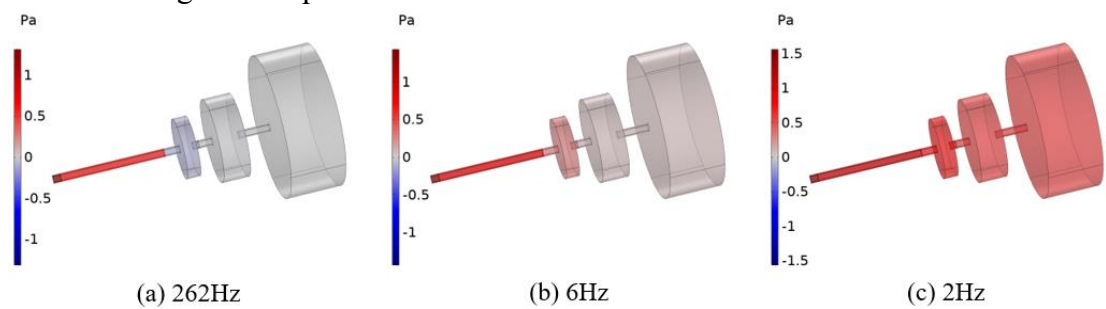

(a) 262Hz  (b) 6Hz  (c) 2Hz

Fig.12 Sound pressure cloud map (after optimization)

Both theoretical predictions and COMSOL results verify the optimization design method. After optimization design, the three-chamber structure can achieve a sound absorption performance of 8.6 octave bands covering super low frequency range. Meanwhile, the structural size is not big, less than $180 \times 100 \times 100 \left( mm^3 \right)$, and can be further reduced through engineering design.

## 5 Conclusions

Based on the MPP theory, through combining the transfer function model of the sound absorption structure and simulated annealing algorithm, an optimization design method is constructed, and the sound absorption structure composed of three MPPs and three expansion chambers is designed and optimized. The main conclusions are as follows.

(1) The theoretical model established by transfer matrix method can accurately predicts the sound absorption performance of the sound structure. For the sound absorption bandwidth, its relative error between theoretical prediction and COMSOL simulation is 3.68%.

(2) The optimization design method can significantly improve the sound absorption performance of the structure. After optimization, the effective sound absorption bandwidth extends from 1324Hz (20-1344Hz) to 1591Hz (4-1595Hz) covering 8.6 octave bands, about 141% of the octave bands before optimization.

(3) The multi-chamber structure composed of MPPs and expansion chambers can achieve surprising low-frequency broadband characteristics and the structural size is not large, overcoming the limitation of poor low-frequency absorption performance of MPP absorbers, which is of great significance to promote the engineering application of MPP absorbers.

**Acknowledgements:**
This project is funded by the National Natural Science Foundation of China (No.51975083, No.51775080, No.11372059, No.11272073)